\documentstyle[aps,epsfig,float]{revtex}

\newcommand{\sphalf}{1.15}

\newcommand{\kimspace}{\edef\baselinestretch{\sphalf}\Large\normalsize}

\begin{document}

\begin{center}
{\bf 
CORRECTIONS TO
FERMI'S GOLDEN RULE IN $\phi \rightarrow K\bar{K}$ DECAYS}
\end{center}

\begin{center}
E. Fischbach$^{(a)}$, A.W. Overhauser$^{(a)}$, and B. Woodahl$^{(b,a)}$
\end{center}

\begin{center}
$^{(a)}$
Physics Department, Purdue University, West Lafayette, IN 47907\\
$^{(b)}$Physics Department, North Dakota State University, Fargo, ND 58105
\end{center}
\vspace{30pt}

\begin{center}
{\bf Abstract}
\end{center}

\begin{quote}
We analyze the decays $\phi \rightarrow K\bar{K}$ utilizing
a formulation of transition rates 
which explicitly exhibits 
corrections to Fermi's Golden Rule.   These corrections arise in
systems in which the phase space and/or matrix element varies rapidly
with energy, as happens in $\phi \rightarrow K\bar{K}$, 
which is just above threshold.    
We show that the theoretical
corrections 
resolve a puzzling $5\sigma$ discrepancy between theory and
experiment for the branching ratio $R = \Gamma (\phi \rightarrow
K^+K^-)/\Gamma(\phi \rightarrow K^0\bar{K}^0)$.
\end{quote}

\pagebreak


One of the most well known results from elementary quantum mechanics
is the formula relating the rate $\Gamma$ for the transition
$|i> \rightarrow |n>$ to the matrix element $V_{ni}(E) \equiv
<n|V|i>$ induced by a time-independent perturbation $V$,
$$ \Gamma(i \rightarrow n) = \frac{2\pi}{\hbar} |V_{ni}(E)|^2 \rho(E),
  \eqno{(1)}
$$
where $E = E_n -E_i$, and $\rho(E) = dN/dE$ is the energy density
of the final states $n$ [1].  This formula is so useful and so widely
applied that Fermi named it ``Golden Rule No. 2" (FGR2) [2].
As almost all derivations of Eq.(1) make clear, FGR2 is
an approximation which is valid in the limit when both
$|V_{ni}(E)|^2$ and $\rho(E)$ are ``slowly varying"
functions of $E$, although the precise meaning of ``slowly
varying" is not always made explicit.
However, it is evident that $|V_{ni}(E)|^2\rho(E)$ will not be
slowly varying in some circumstances, for example, when the
initial state $|i>$ is just above the threshold for decay into
the final state $|n>$.  This is the case for the decays
$\phi \rightarrow K\bar{K}$ (either $\phi \rightarrow K^+K^-$
or $\phi \rightarrow K^0\bar{K}^{0})$, where $m_\phi = 1019.417(14)MeV$,
while $m_{K^+} = 493.677(16)MeV$, and $m_{K^0} = 497.672(31) MeV$ [3].
Quark model diagrams for these decays are shown in Fig. 1.
In principle the corrections to FGR2 in such decays could be
large enough to lead to detectable effects, and in what follows
we show that this is in fact the case.  More interestingly,
we demonstrate explicitly that the correction to FGR2 arising
from the rapid variation of $|V_{ni}(E)|^2\rho(E)$ 
with $E$ 
resolves a puzzling discrepancy [5] between theory and
experiment for the ratio
$R = \Gamma(\phi \rightarrow K^+K^-)/\Gamma(\phi \rightarrow
K^0\bar{K}^0)$. 

An explicit expression for the corrections to FGR2 can be conveniently
derived by endowing the initial decaying state at the outset with
a lifetime $\tau = 1/\Gamma$, and then solving self-consistently
for $\Gamma$.  Consider an initial state $|i,t_0>$ which evolves
into the state $|i,t>$ at a later time $t$ under the influence
of the time evolution operator $U(t,t_0)$.  The state $|i,t>$ can be
expanded in terms of a complete set of eigenfunctions  $|n>$
of the unperturbed Hamiltonian,
$$  \everymath={\displaystyle}
  \begin{array}{rcl}
   |i,t> &=& \sum_n |n><n|U(t,t_0)|i,t_0> \equiv \sum_n c_n(t)|n>,\\
    &~& \\
   c_n(t) &=& \frac{-i}{\hbar} \int_{t_0}^t dt^\prime
      e^{i\omega_{ni}t^\prime} e^{-\Gamma t^\prime/2} V_{ni},
   \end{array}
  \eqno{(2)}
$$
where $\hbar \omega_{ni} = E_n-E_i$, and
$E_i(E_n)$ is the unperturbed initial (final) energy.
Without loss of generality we can set $t_0 = 0$, 
the instant at which the $\phi$ is produced. 
The quantity of interest is $c_n{(\infty)}$ which is given by
$$  c_n({\infty}) =  \frac{-i}{\hbar} \int_0^\infty
    dt^\prime e^{(i\omega_{ni}-\Gamma/2)t^\prime} V_{ni}
     = \frac{i}{\hbar} V_{ni} \frac{1}{i\omega_{ni} - \Gamma/2},
    \eqno{(3)}
$$
where $V_{ni}$ is independent of time.
We next impose the unitarity constraint on $c_n({\infty})$, namely
$\sum_n|c_n({\displaystyle{\infty}})|^2 =1 $.  After the sum is converted into an
integral in the usual manner, $\sum_{n} \rightarrow \int dE\rho(E)$,
the unitarity constraint assumes the form
$$  1 = \int_{-\infty}^\infty dE\rho(E) \frac{|V(E)|^2}{E^2+(\hbar\Gamma/2)^2}~~~,
  \eqno{(4)}
$$
where we have set $\hbar \omega_{ni} \rightarrow E$, and
$V_{ni} \rightarrow V(E)$.  The denominator in Eq.(4) is
rapidly varying in the vicinity of $E \equiv E_0 \cong 0$,
which corresponds to an energy-conserving transition.  Thus if
we invoke the assumption that $\rho(E)|V(E)|^2$ is slowly varying
{\it with respect to the denominator} [$E^2+(\hbar\Gamma/2)^2]$,
then the unitarity constraint in Eq.(4) yields
$$  1 \cong \rho(E_0)|V(E_0)|^2 \int_{-\infty}^\infty dE
    \frac{1}{E^2+(\hbar\Gamma/2)^2}  = \rho(E_0)|V(E_0)|^2 \cdot
    (2\pi/\hbar \Gamma).
   \eqno{(5)}
$$
Solving Eq.(5) for $\Gamma$ we are led immediately to the standard Fermi
Golden Rule in Eq.(1).  
Moreover, the Golden Rule integral technique (GRIT) 
embodied in Eq.(4)
gives a specific formula for the corrections to FGR2 in Eq.(1)
for processes in which $\rho(E)$ and/or $|V(E)|^2$ varies significantly 
with energy.  One can further elucidate the approximation
being made in going from Eq.(4) to Eq.(5) by invoking
the identity
$$  \lim_{\alpha \rightarrow 0} \frac{1}{E^2 + \alpha^2} =
    \frac{\pi}{\alpha} \delta(E) ,
   \eqno{(6)}
$$
where $\alpha = \hbar\Gamma/2$.  Combining Eqs.(4) and (6)
leads immediately to Eq.(5) in the limit $\Gamma \rightarrow 0$.
However, for $\Gamma \not= 0$ $\delta(E)$ is replaced by
the (broader) Lorentzian in Eq.(4) which introduces
contributions from $\rho(E)|V(E)|^2$ in which $E \not= 0$.
As we now demonstrate, these additional contributions are
quantitatively different for $\phi \rightarrow K^+ K^-$
and $\phi \rightarrow K^0\bar{K}^0$, and their inclusion
serves to resolve a $5\sigma$
discrepancy between the theoretical
and experimental values [4,5] of $R = \Gamma(\phi \rightarrow K^+K^-)/
\Gamma(\phi \rightarrow K^0\bar{K}^0$).

The decay $\phi \rightarrow K^+ K^-$ is induced by the 
Lagrangian density
$$  {\cal{L}}(x) = ig_+ \phi^\mu (x) [K^+(x)\partial_\mu K^-(x)-K^-(x)
    \partial_\mu K^+(x)],
  \eqno{(7)}
$$
where $g_+$ is the appropriate coupling constant, and $K^+(x)$
annihilates $K^+$, etc.  A similar expression characterizes
$\phi \rightarrow K^0 \bar{K}^0$, which is proportional
to the coupling constant $g_0$, with $g_0 = g_+$ in  the 
limit of exact SU(2) symmetry.  Bramon, {\it et al.} [5]
have considered the effects of radiative corrections,
and we will return to their results
below.  Using FGR2 as given in Eq.(1) the decay rate
$\Gamma(\phi \rightarrow K^+ K^-)$ obtained from Eq.(7) is
given by
(setting $\hbar = c = 1$ hereafter),
$$  \Gamma(\phi \rightarrow K^+K^-) = \frac{2}{3}
    \left(\frac{g_+^2}{4\pi}\right) \frac{|\vec{k}|^3}{m_\phi^2},
  \eqno{(8)}
$$
where $|\vec{k}|= (1/2) (m_\phi^2 - 4m_{K^+}^2)^{1/2}$
is the magnitude of the $K^+$ 3-momentum in the $\phi$ rest frame.
The factor of $|\vec{k}|^3$ can be understood as follows:
Since the kaons are spinless, whereas $\phi$ is a vector particle,
angular momentum conservation demands that $K$ and $\bar{K}$
be emitted in a relative $P-$wave, which is 
consistent with
the derivative coupling in Eq.(7).  Hence $|$$<V_{ni}>$$|^2$ contributes
a factor of $|\vec{k}|^2$, while the phase space
contributes an additional factor $dN/dE \propto |\vec{k}|$.
Combining Eq.(8) with the corresponding expression for
$\phi \rightarrow K^0\bar{K}^0$ we find using FGR2,
$$  R_{th} \equiv \frac{\Gamma (\phi \rightarrow K^+K^-)}
                       {\Gamma (\phi \rightarrow K^0\bar{K}^0)}
    \Biggr\vert_{th} \equiv 
          \left(\frac{g_+^2}{g_0^2}\right) R_{FGR2} =
        \left(\frac{g_+^2}{g_0^2}\right) 
                \left(\frac{1 -4 \mu_+^2}
                                         {1-4\mu_0^2}\right)^{3/2},
   \eqno{(9)}
$$
where $\mu_+ = m_{K^+}/m_\phi$ and $\mu_0 = m_{K^0}/m_\phi$.  Inserting
the previously quoted values of $m_{K^+}, m_{K^0}$, and
$m_\phi$ into Eq.(9), and assuming $g_0 = g_+$, we find
$R_{th} = 1.528$, to be compared with the experimental
value [4]

$$  R_{exp} = 1.456 \pm 0.033  ~~~.
   \eqno{(10)}
$$

Bramon, {\it et al.} [5] have evaluated various corrections
to $R_{th}$ in an effort to bring $R_{th}$ and $R_{exp}$ into
agreement.  Most significant among these are electromagnetic
radiative corrections which affect $\phi \rightarrow K^+K^-$
but not $\phi \rightarrow K^0 \bar{K}^0$.
These authors find that the radiative correction factor, 
$\eta = 1.042$, {\it increases}
$R_{th}$ to 1.59. 
Bramon {\it et al.} have also studied the effects of
SU(2) symmetry breaking on the ratio $g_+/g_0$, which arise
via quark mass differences.  The $\phi$ wavefunction is
pure $s\bar{s}$, and hence the $K^+K^-$ $(K^0\bar{K}^0)$
final state requires the creation of an additional
$u\bar{u}(d\bar{d})$ pair (see Fig. 1).  Since the $u\bar{u}$ pair is
lighter than $d\bar{d}$, one expects this effect to enhance
$\phi \rightarrow K^+K^-$ relative to $\phi \rightarrow K^0\bar{K}^0$,
thus further widening the discrepancy between theory and
experiment.  A detailed analysis by Bramon {\it et al.} [5]
finds $g_+/g_0 \cong 1.01$, which agrees with the intuitive
expectation that this correction also works in the wrong
direction.  With this correction included $R_{th}$ becomes 1.62. 
Other effects considered by these authors,
such as the inclusion of electromagnetic form factors in calculating
radiative corrections, and final-state rescattering
effects, are negligible.  We are thus left with a puzzling
$5\sigma$ discrepancy between $R_{th} = 1.62$ and
$R_{exp} = 1.456~(33)$.

We proceed to demonstrate that this discrepancy can be resolved 
by incorporating the corrections to FGR2 that arise from the Golden
Rule integral technique.  Combining Eqs.(4) and (8),
and introducing the notation $z = E/m_\phi$,
$a = \Gamma/2m_\phi$, 
$(\Gamma = 4.458(32)$MeV)
we express the unitarity
constraint for $\phi$-decays in the form,
$$  \everymath={\displaystyle} 
\begin{array}{rcl}
   1 &=& \frac{1}{3\pi} \left(\frac{g_+^2}{4\pi}\right)
      \int \frac{dz}{(1+z)} 
    \frac{1}{z^2+a^2} \left[\frac{1}{4} (1+z)^2 - \mu_+^2\right]^{3/2}\\
    & &\\
   &+& \frac{1}{3\pi} \left(\frac{g_0^2}{4\pi}\right) \int
       \frac{dz}{(1+z)} \frac{1}{z^2 + a^2} \left[\frac{1}{4} (1+z)^2 - 
     \mu_0^2\right]^{3/2}~~ + ... ~~~~.
   \end{array}
   \eqno{(11)}
$$
The two terms exhibited in Eq.(11) are, respectively, the contributions
from the $K^+ K^-$ and $K^0 \bar{K}^0$ states, and $...$ denotes
contributions to the unitarity integral from other channels
(such as $\rho\pi)$ which can be ignored for present purposes.
We emphasize that the functional form of the expressions in
square-brackets in Eq.(11) is determined by the kinematics of
$\phi \rightarrow K\bar{K}$, specifically by the relation between
$|\vec{k}|$ and $E$ given in Eq.(12) below.  
The density of final states is readily found to be proportional
to $(1+z)k$,
and each of the bosonic normalization coefficients $(2E_K)^{-1/2}$
contributes a factor $(1+z)^{-1}$, resulting in an overall
factor $(1+z)^{-1}$.
Eq.(12) also fixes the lower limit of integrations in
Eq.(11) as we discuss below.
$R_{th}$ is given by the ratio of the two terms in (11),
which in the narrow resonance $(a \rightarrow 0)$ limit gives the
standard result in Eq.(9).  To specify the integration limits
we note from Eq.(4) that since $E = \hbar \omega_{ni}$ is the energy
difference between the final state $|n>$ and the initial state
$|i>$, we can
write for $\phi \rightarrow K^+ K^-$
in the $\phi$ rest frame,
$$  E = 2 \sqrt{|\vec{k}|^2 + m_{K^+}^2} - m_\phi .
    \eqno{(12)}
$$
The lower limit on $E$ evidently corresponds to $|\vec{k}| = 0$,
and gives $E_{min} = 2m_{K^+} - m_\phi$.
Accordingly in Eq.(11),  $z_{min} = (2\mu_+ - 1)$ for
$\phi \rightarrow K^+ K^-$, and $z_{min} =
(2\mu_0 -1)$ for $\phi \rightarrow {K}^0 \bar{K}^0$.  The upper limit
on $|\vec{k}|$ (and hence $z$) extends to infinity. 
This limit leads to  divergent integrals
in Eq. (11), so the unitarity constraint (here as elsewhere) 
requires modification of the high-energy behavior of the
$\phi K\bar{K}$ amplitudes.  
This can be achieved 
by incorporating
a phenomenological form factor,
$$  F(|\vec{k}|^2) = \frac{M^2}{M^2+|\vec{k}|^2} ,
   \eqno{(13)}
$$
multiplying the $\phi K\bar{K}$ amplitude.  This 
form factor introduces an asymptotic $1/z^4$ dependence
(after the $\phi K\bar{K}$ amplitude is squared); so convergence is
assured.  The energy scale, $M$, is related to the
confinement size of the hadrons involved (compared to $1/k$),
and is typically of order $\sim 1~GeV$ [6].
We have calculated $R_{GRIT}$ numerically as a function of $M$,
and combined those results with the radiative correction
factor $\eta = 1.042$ and the SU(2) correction
$(g_+^2/g_0^2) = 1.02$ to obtain $R_{th}$,
$$  R_{th} = \frac{g_+^2}{g_0^2} \eta R_{GRIT} ~~~.
   \eqno{(14)}
$$
A plot of
$R_{th}$ as a function of $M$ is shown in Fig. 2,
along with the $1\sigma$ experimental result from Eq.(10)
which is indicated by the dashed horizontal line.
We see from this figure that $R_{th}$ is relatively insensitive
to the choice of $M$, and that for $M\stackrel{>}{\sim} 0.8~GeV$
$R_{th}$ falls within the $1\sigma$ experimental bounds.
For the nominal value $M=1~GeV$ we find $R_{th} = 1.48$ compared
to the experimental value $R_{exp} = 1.456~(33)$.

The reduction in $R$ relative to its value derived from FGR2
can be understood by considering the integrand of the $K^0\bar{K}^0$
integral in Eq.(11), shown in Fig. 3.  The factor multiplying the
Lorentzian denominator is asymmetric about $z=0$.  The contribution
from $z > 0$ significantly exceeds the result obtained if this factor
is replaced by its $z=0$ value.  The proportionate increase is greater
for the $K^0\bar{K}^0$ decay than for the $K^+K^-$ decay because
$\phi \rightarrow K^0\bar{K}^0$ is closer to threshold,
so $R$ becomes smaller than the value in Eq.(9).

Although the discrepancy between the theoretical values of
$R_{GRIT}$ and $R_{FGR2}$ is $\sim 9\%$, the corrections to
the individual partial decay rates are larger.  $\Gamma_{GRIT}/\Gamma_{FGR2}$
is shown as a function of $M$ in Fig. 4 for both $K^+K^-$
and $K^0\bar{K}^0$ channels.  It seems evident that
corrections to FGR2, similar to those considered here
(but not necessarily so dramatic) can be anticipated in
other decays.

We wish to thank G. Pancheri, K. Gottfried, D. Gottlieb, M. Haugan, H. Rubin, and
S.J. Tu for helpful discussions.  One of the authors (E.F.) wishes
to acknowledge the support of the U.S. Department of Energy
under contract No. DE-AC02-76ER01428.

\pagebreak
  
\kimspace

\begin{center}
{\bf REFERENCES}
\end{center}

\begin{description}
\item[[1]]  L.I. Schiff, {\it Quantum Mechanics},
2nd ed., (McGraw-Hill, New York, 1955) p. 199;
W. Heitler, {\it The Quantum Theory of Radiation},
2nd ed., (Oxford Univ. Press, London, 1944) p. 113.

\item[[2]] E. Fermi, {\it Nuclear Physics},
Revised Edition, (Univ. of Chicago Press, 1950)
pp. 141-142; 214; see also, {\it Elementary Particles}
(Yale Univ. Press, New Haven, 1951) p. 30.
The first derivation of $\Gamma$ is due to:  P.A.M. Dirac,
Proc. Roy. Soc. (London), A114, 243 (1927), Eq. 32.
(Fermi elevated Dirac's formula to the status of ``Golden Rule".)

\item[[3]] Particle Data Group, Eur. Phys. J. C15
(2000) 1, p. 420, 494, 509.

\item[[4]] ibid., p. 421.

\item[[5]] A. Bramon, R. Escribano, J.L. Lucio M., and
G. Pancheri, Phys. Lett. B486 (2000) 406.
The conclusions of this reference have been confirmed by:
M. Benayoun and H.B. O'Connell, nucl-th 0107047, p. 26.

\item[[6]] M.L. Perl, {\it High Energy Hadron Physics}
(Wiley, New York, 1974), p. 453.
\end{description}

\newpage
\begin{figure}
{\centering \resizebox*{4.0in}{!}{\includegraphics{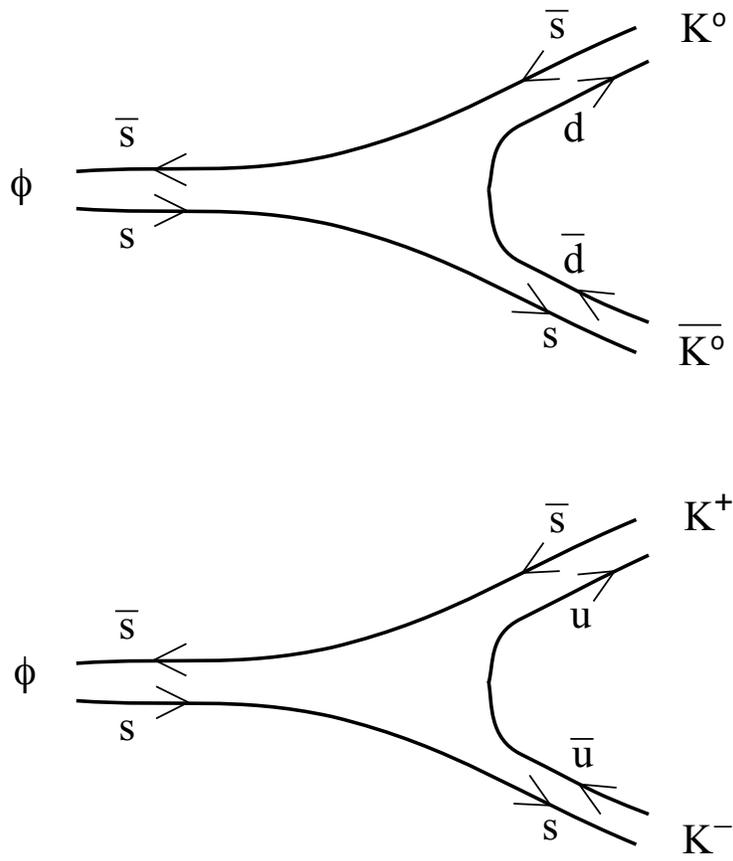}} \par}
\caption{Quark model diagrams for $\phi \rightarrow K^+K^-$ and $\phi \rightarrow K^0\bar{K}^0$}
\end{figure}

\newpage
\begin{figure}
{\centering \resizebox*{4.0in}{!}{\includegraphics{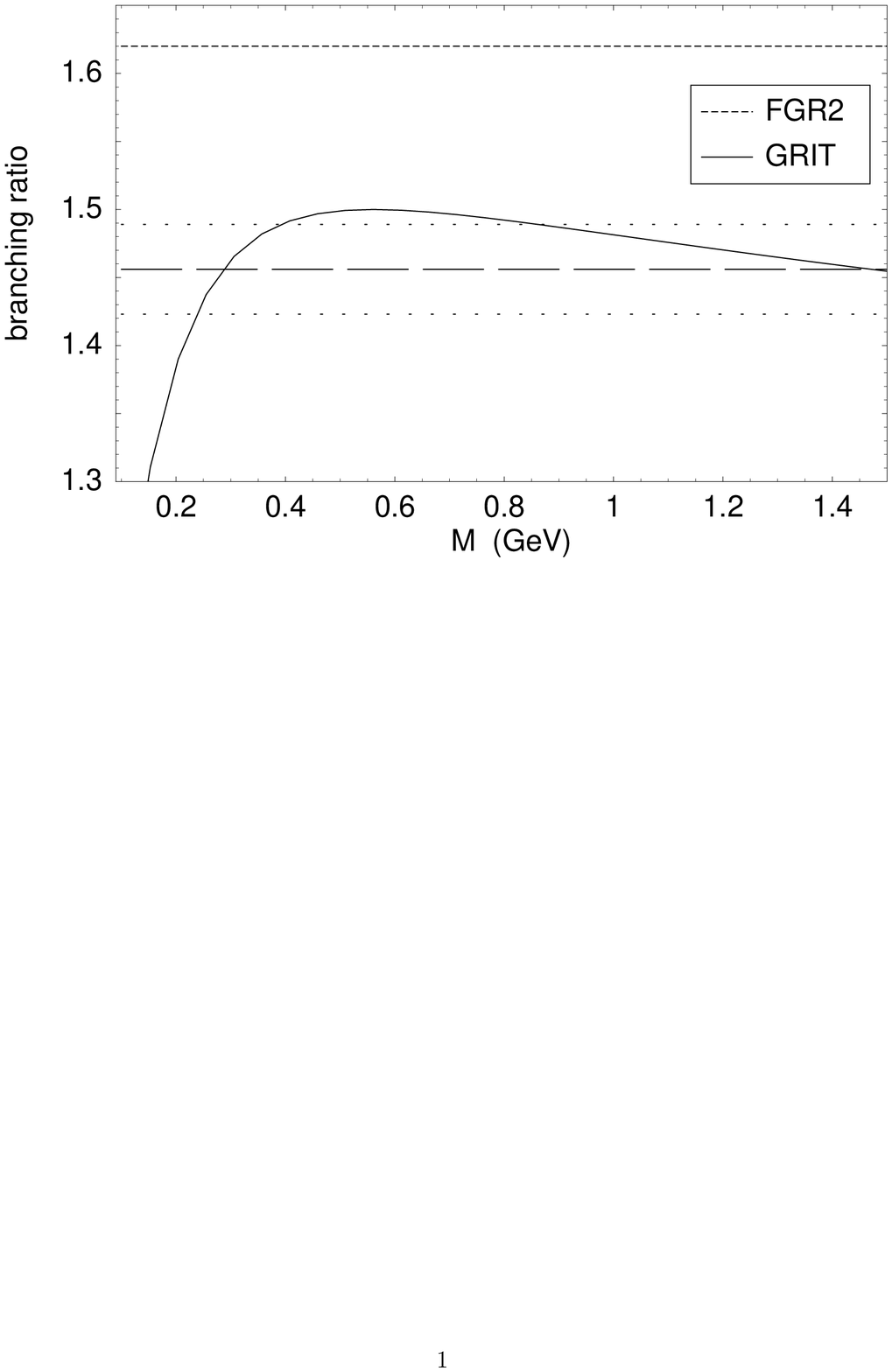}} \par}
\caption{The solid curve gives the
theoretical branching ratio versus $M$, the energy scale
in Eq.(13), where $R_{th} = (g_+/g_0)^2 \eta R_{GRIT}\approx 1.063 R_{GRIT}$.
The experimental ratio in Eq.(10) is given by the long-dashed line,
and the $1\sigma$ bounds fall within the dotted lines.
The short-dashed line is the result predicted by FGR2.}
\end{figure}

\newpage
\begin{figure}
{\centering \resizebox*{4.0in}{!}{\includegraphics{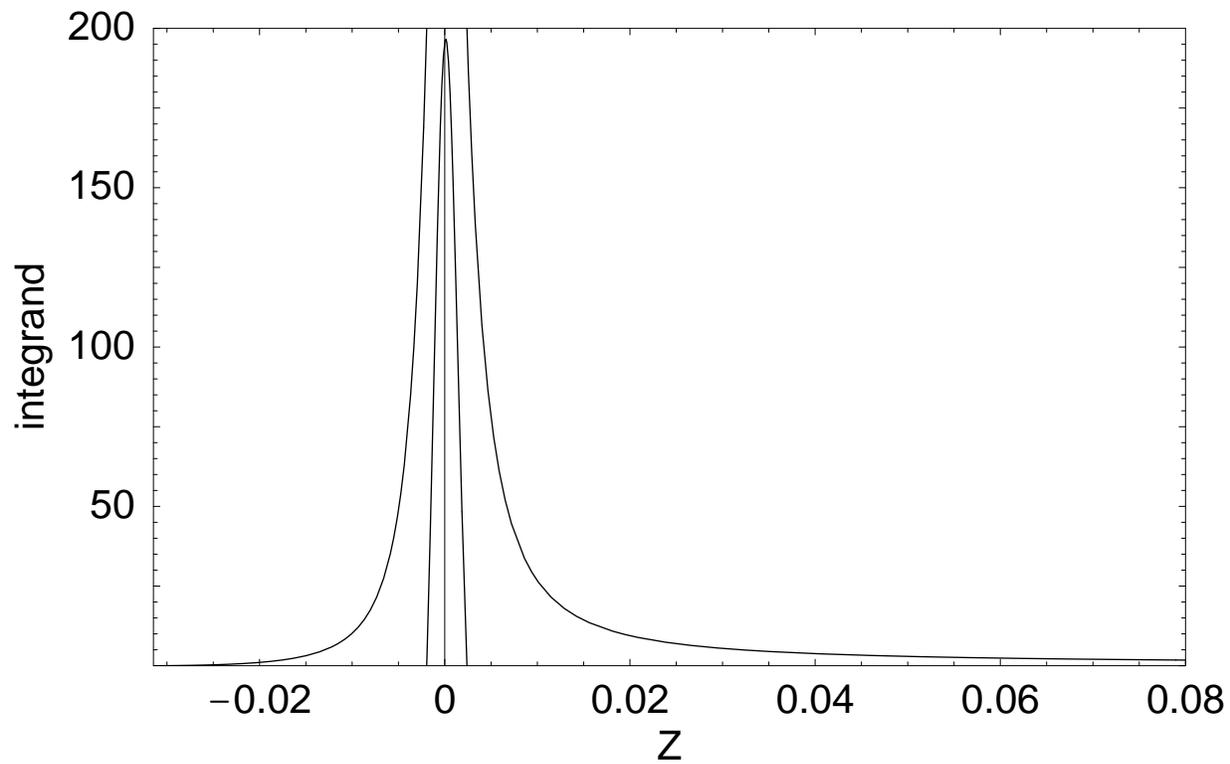}} \par}
\caption{The $K^0\bar{K}^0$ integrand in Eq.(11) multiplied
by the square of the form factor, Eq.(13).
For the central peak, the ordinate is obtained by adding 200 to the
value shown.}
\end{figure}

\newpage
\begin{figure}
{\centering \resizebox*{4.0in}{!}{\includegraphics{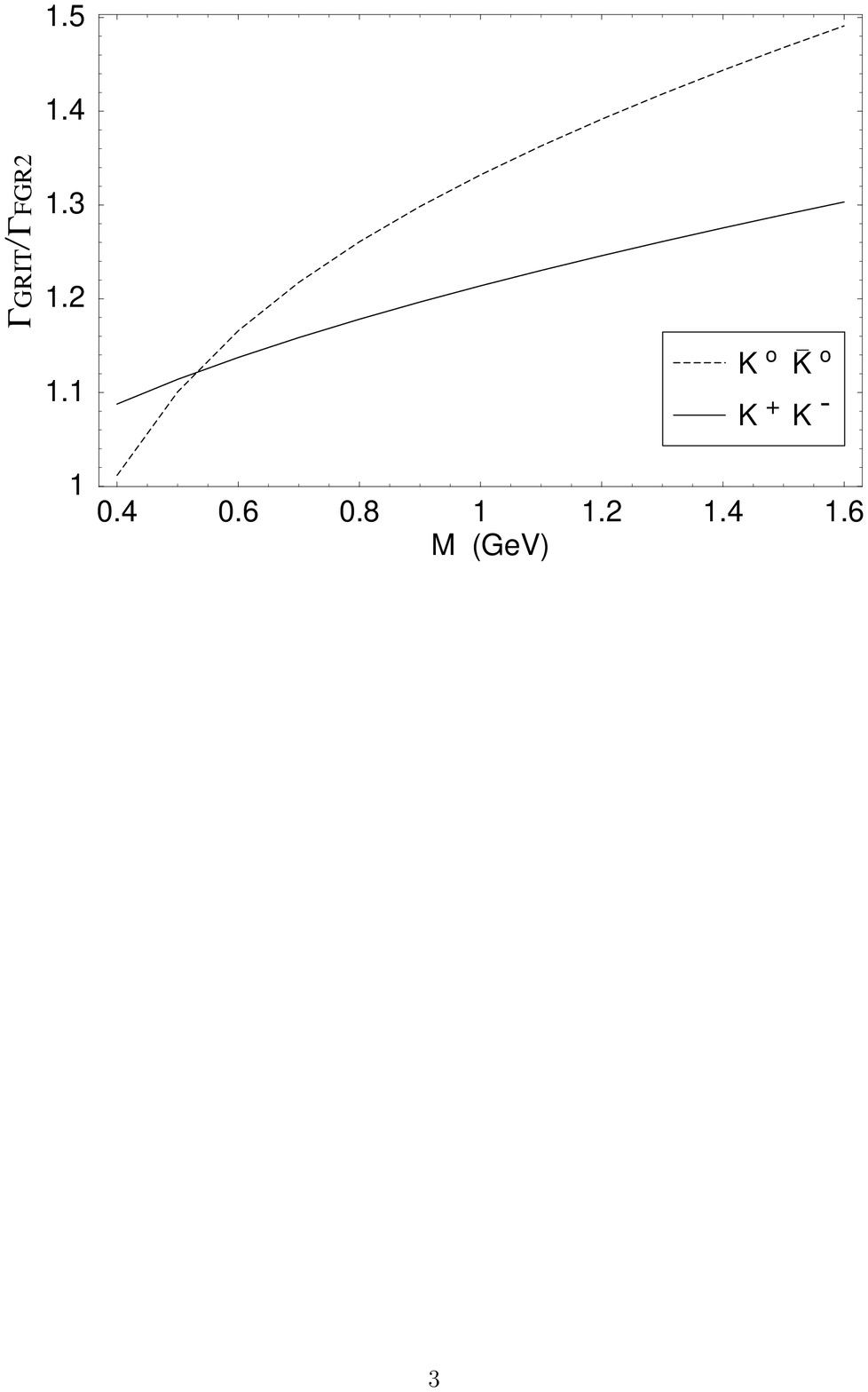}} \par}
\caption{$\Gamma_{GRIT}/\Gamma_{FGR2}$ versus $M$ for both
the $K^+K^-$ (solid curve) and $K^0\bar{K}^0$ (dashed curve) channels.}
\end{figure}

\end{document}